\documentclass[12pt]{article}
\usepackage{lmodern}
\usepackage{xr}
\usepackage[margin=1in]{geometry}
\usepackage[utf8]{inputenc}
\usepackage{graphicx} 
\usepackage[square,numbers]{natbib}
\usepackage{amsmath} 
\usepackage{newtxtext} 
\usepackage{hyperref}
\usepackage[table,x11names]{xcolor}
\usepackage{authblk}
\usepackage{booktabs}
\usepackage{threeparttable}
\usepackage{siunitx}
\usepackage{dcolumn}
\usepackage{rotating}
\usepackage{setspace}
\usepackage[font=small,labelfont=bf]{caption}
\usepackage{subcaption}

\usepackage{float} 

\newcommand{\AI}{\mathrm{AI}}
\newcommand{\RCA}{\mathrm{RCA}}
\newcommand{\PCI}{\mathrm{PCI}}
\newcommand{\AECI}{\mathrm{AECI}}
\newcommand{\AIAP}{\mathrm{AIAP}}
\newcommand{\AECP}{\mathrm{AECP}}
\graphicspath{{figures/}}

\title{\textbf{\LARGE Mapping AI Economic Complexity}\\
\vspace{0.3cm}
}

\author[1]{Daeun Moon}
\author[1,2,3]{Yeokyung Hwang}
\author[1]{Junseok Hwang}
\author[2,3*]{Dawoon Jeong}

\affil[1]{Technology Management, Economics and Policy Program (TEMEP), Seoul National University, Seoul, Republic of Korea}
\affil[2]{Department of Sociology, University of Chicago, Chicago, IL, USA}
\affil[3]{Knowledge Lab, University of Chicago, Chicago, IL, USA}
\affil[*]{Correspondence: Dawoon Jeong (jdwoon0523@uchicago.edu)}
\date{\today}

\begin{document}
\maketitle
\begin{abstract}
Green economic complexity provides a generalizable framework for examining countries' productive capabilities in a defined product set. We apply this framework to AI-enabling goods within the full product space, linking current specialization with adjacent diversification opportunities. Using BACI exports for 2007--2023 and 103 AI-enabling goods, we measure complexity-weighted specialization (AECI), product-level adjacent opportunities (AIAP), and average complexity-weighted relatedness of remaining candidates (AECP). In 2023, Japan leads AECI, while China leads AECP; portfolio breadth accounts for much of the variation in raw AECI. Initial raw potential is positively associated with subsequent changes in the AI-enabling export share, but its associations with changes in AECI and specialization counts are not statistically significant at the 5\% level. Our contribution is a trade-based assessment of AI-enabling productive capabilities and related opportunities. The results and public dashboard provide a preliminary complement to publication and patent indicators, not a comprehensive measure of national AI performance or a validated forecast of diversification.
\end{abstract}

\section{Introduction}
\label{sec:introduction}
National AI capability encompasses not only scientific and inventive output but also the productive structures supporting AI's material infrastructure. Publication and patent counts do not, by themselves, reveal which countries competitively export the equipment, components, and materials enabling AI systems. The WTO's AI-enabling goods classification provides a basis for examining this production dimension \citep{wto2025}. We ask which productive strengths countries reveal through these exports and how remaining AI-enabling activities relate to their existing specializations.

Economic complexity examines country--product specialization as indirect evidence of productive capabilities \citep{hidalgo2009}. Product-space relatedness connects existing specializations with potential new activities \citep{hidalgo2007}. Together, these approaches distinguish the scale of exports from the breadth and complexity of specialized portfolios, and situate remaining activities relative to those portfolios.

Research on green economic complexity shows how this perspective can be focused on a defined product set. \citet{mealy2022} distinguish current green capabilities, adjacent opportunities, and diversification potential, and explicitly note that their complexity index can be applied to any subset of products. This provides a generalizable framework: the selected goods define the domain of interest, while the wider product space supplies the information for assessing capabilities and related opportunities.

We apply this framework to AI-enabling goods. Our contribution is to map countries' revealed AI-enabling productive capabilities and adjacent diversification opportunities within the full product space. AECI characterizes current specialization by its breadth and complexity, AIAP retains product-level candidates, and AECP summarizes their average complexity-weighted relatedness. We first compare current portfolios and remaining opportunities, then examine preliminary conditional associations between initial raw potential and later changes in export composition and specialization. The emphasis is on productive specialization and structurally related opportunities, not a comprehensive ranking of national AI performance.

\section{Data \& Framework}
\label{sec:framework}
\subsection{Country--product exports and AI-enabling goods}
We use BACI HS 2007, 2007--2023 \citep{gaulier2010}, aggregating exports over foreign destinations. Annual matrices $X_{cpt}$ cover 228 economies and 5,047 HS6 products in current thousand US dollars. The fixed target set $A$ contains 103 goods (52 HS4 headings) from the project's HS 2012 (H4) version of the WTO AI-enabling list \citep{wto2025}; all identifiers occur in the HS 2007 matrices. The list is applied retrospectively and does not identify exclusively AI end uses. All products supply the information used to evaluate complexity and relatedness.

\subsection{Product specialization and AECI}
For country $c$, product $p$, and year $t$, revealed comparative advantage and the AI indices' specialization matrix are
\begin{equation}
\RCA_{cpt}=\frac{X_{cpt}/\sum_j X_{cjt}}{\sum_i X_{ipt}/\sum_{i,j} X_{ijt}},
\qquad M_{cpt}=\mathbf{1}\{\RCA_{cpt}>1\}.
\label{eq:rca}
\end{equation}
All sums in Eq.~\eqref{eq:rca} use the full annual trade matrix. Zero-denominator RCA is stored as zero; $M=0$ denotes no revealed specialization, not necessarily no exports. Define the AI-enabling export share as $s_{ct}^{\AI}=\sum_{p\in A}X_{cpt}/\sum_jX_{cjt}$. For comparison, \emph{bundle RCA} divides this share by the corresponding share of world exports. It measures relative specialization in the bundle, rather than the breadth of a country's AI-enabling portfolio.

The Product Complexity Index (PCI) is taken from the full-product-space pipeline. Suppressing the year index, on positive-degree rows and columns write $B_{cp}=\mathbf{1}\{\RCA_{cp}\geq1\}$ and let $D_c,D_p$ be its diagonal degree matrices. The product score uses the second right singular vector of $D_c^{-1/2}BD_p^{-1/2}$, rescaled by $D_p^{-1/2}$, oriented negatively with ubiquity, and standardized across products. Thus PCI uses $\RCA\geq1$, whereas Eq.~\eqref{eq:rca} uses the strict rule. We retain and disclose this implementation difference. Let $P_t^*$ contain all products with finite PCI and define
\begin{equation}
q_{pt}=\frac{\PCI_{pt}-\min_{j\in P_t^*}\PCI_{jt}}
{\max_{j\in P_t^*}\PCI_{jt}-\min_{j\in P_t^*}\PCI_{jt}},
\quad F_{ct}=\sum_{p\in A}M_{cpt}q_{pt},
\quad \AECI_{ct}=\frac{F_{ct}-\bar F_t}{s_{F,t}}.
\label{eq:aeci}
\end{equation}
Normalization of $q$ is across \emph{all} finite-PCI products, not only AI goods. The AI Economic Complexity Index (AECI) uses binary membership weights and standardizes the \emph{sum} of normalized complexity across specialized AI goods \citep{mealy2022}. Its decomposition is $F_{ct}=D_{ct}^{\AI}\bar q_{ct}^{\AI}$, where $D_{ct}^{\AI}=\sum_{p\in A}M_{cpt}$ and $\bar q_{ct}^{\AI}$ is mean complexity among those goods when $D_{ct}^{\AI}>0$.

\subsection{AIAP and AECP}
Let $u_{pt}=\sum_c M_{cpt}$ be product ubiquity. Following the product-space approach \citep{hidalgo2007}, proximity and country-specific relatedness density are
\begin{equation}
\phi_{jpt}=\frac{\sum_c M_{cjt}M_{cpt}}{\max(u_{jt},u_{pt})}\quad(j\ne p),
\qquad \rho_{cpt}=\frac{\sum_{j\ne p}M_{cjt}\phi_{jpt}}{\sum_{j\ne p}\phi_{jpt}}.
\label{eq:density}
\end{equation}
Self-links are excluded and zero-denominator proximities are set to zero. Density measures relatedness to the country's \emph{entire} specialized portfolio, including non-AI goods; it is not a geometric distance in a plotted network. Undefined density is left missing.

The AI Adjacent Possible (AIAP) is a set of candidate products with density and complexity coordinates, not a scalar score:
\begin{equation}
\AIAP_{ct}=\big\{(p,\rho_{cpt},q_{pt}):p\in A,\ M_{cpt}=0\big\}.
\label{eq:aiap}
\end{equation}
No density cutoff is imposed. Let $V_{ct}$ index candidate products with finite density and complexity. The raw potential $G$ and its standardized AI Economic Complexity Potential (AECP) are
\begin{equation}
G_{ct}=\frac{1}{|V_{ct}|}\sum_{p\in V_{ct}}\rho_{cpt}q_{pt},
\qquad \AECP_{ct}=\frac{G_{ct}-\bar G_t}{s_{G,t}}.
\label{eq:aecp}
\end{equation}
AECP summarizes the \emph{average} complexity-weighted relatedness of remaining opportunities, rather than their total \citep{mealy2022}. It is undefined when $V_{ct}$ is empty. In the reported results, all AI products have finite complexity and all candidate densities are finite, so $|V_{ct}|=103-D_{ct}^{\AI}$.

\subsection{Reporting population and interpretation}
For AECI and AECP, annual means and sample standard deviations ($n-1$ denominator) use countries with at least US\$1 billion of total exports and at least one specialized product. This reporting and normalization rule does not screen the matrices used for RCA or relatedness. The valid sample contains 140--163 economies annually, including 162 in 2023. Scores are relative to each year's reference population and complexity scale; they are not absolute units of AI capability. Numerical consistency checks reconstruct listed-product RCA, $q$, $F$, $G$, and their standardizations from the derived tables. They do not independently verify the underlying BACI ingestion, full-space PCI or proximity estimation, or the substantive validity of the measures.

\subsection{Descriptive comparisons and longitudinal associations}
\label{sec:empirical_design}
Whole-space ECI is the country-level complexity score from the full-product-space pipeline; it is a comparator and control, not a component of AECI. We compare AECI with its components, ECI, bundle RCA, and AECP. Country-level correlations use paired observations within the eligible annual sample. For a separate descriptive HS6 network, proximity is computed using $B_{cp}=\mathbf{1}\{\RCA_{cp}\geq1\}$ and averaged over 2007--2023. The fixed backbone combines a maximum spanning tree with the union of up to four strongest links selected per node whose mean proximity is at least 0.60. Louvain communities summarize this backbone. Neither the pooled backbone nor its layout or communities enters the annual AECP calculation, which uses the full annual proximity matrix based on the strict specialization rule in Eq.~\eqref{eq:rca}.

For the longitudinal comparison, the outcomes $Y_{ct}$ are the AI-enabling export share $s_{ct}^{\AI}$, standardized AECI, and specialization count $D_{ct}^{\AI}$. We compare baseline (2007--2011) and follow-up (2019--2023) means using
\begin{equation}
\Delta\bar Y_c
=\alpha+\beta\bar G_{c0}
+\gamma\overline{\mathrm{ECI}}_{c0}
+\delta\overline{\log y}_{c0}
+\eta\bar Y_{c0}+\varepsilon_c,
\label{eq:change_regression}
\end{equation}
where $\Delta\bar Y_c=\bar Y_{c1}-\bar Y_{c0}$ and subscripts 0 and 1 denote the two periods. The regressor $\bar G_{c0}$ is mean baseline \emph{raw} potential, not standardized AECP. Annual real GDP per capita $y_{ct}$ is output-side real GDP divided by population (\texttt{rgdpo/pop}) in Penn World Table 11.0 \citep{feenstra2015}; $\overline{\log y}_{c0}$ is the mean of annual log income, not the log of mean income. Baseline means use eligible years with jointly observed outcomes and regressors; follow-up means use eligible years with observed outcomes. Each regression requires at least three usable years per period and uses ordinary least squares with HC1 heteroskedasticity-robust standard errors. These are exploratory, in-sample associations rather than causal or out-of-sample tests.

\section{Preliminary Results}
\label{sec:results}

\subsection{Scale, complexity, and country portfolios}
\label{sec:portfolios}
AI-enabling goods account for 11.31\% of world exports in 2007 and 12.44\% in 2023, with nominal exports rising from US\$1.52 trillion to US\$2.82 trillion. In 2023, the five largest exporters account for 55.2\% of these exports. Among the 5,026 products with finite PCI in 2023, the 103 AI-enabling goods have a mean PCI of 0.837, compared with approximately zero for the full product set; 58.3\% lie in the highest complexity quartile. These values describe trade in the listed categories, not exports used exclusively for AI.

\begin{table}[H]
\centering
\begin{threeparttable}
\caption{Leading AECI economies and selected comparators, 2023}
\label{tab:country_portfolios}
\small
\setlength{\tabcolsep}{6pt}
\begin{tabular*}{\textwidth}{@{\extracolsep{\fill}}lrrrrr@{}}
\toprule
Economy
& $D^{\AI}$
& $\bar q^{\AI}$
& AECI
& $|V|$
& AECP \\
\midrule
Japan         & 64 & 0.691 & 3.996 & 39 & 1.516 \\
China         & 66 & 0.623 & 3.666 & 37 & 4.144 \\
Germany       & 61 & 0.654 & 3.535 & 42 & 3.382 \\
United States & 53 & 0.656 & 2.991 & 50 & 1.918 \\
Taiwan        & 52 & 0.666 & 2.977 & 51 & 0.836 \\
South Korea   & 40 & 0.692 & 2.241 & 63 & 0.802 \\
Italy         & 35 & 0.629 & 1.643 & 68 & 3.442 \\
\bottomrule
\end{tabular*}
\begin{tablenotes}[flushleft]
\footnotesize
\item Notes: The first five rows are the five highest-AECI economies in 2023; South Korea and Italy are included as selected comparators. $D^{\AI}$ is the number of AI-enabling goods in which an economy has revealed specialization, and $\bar q^{\AI}$ is their mean normalized product complexity. $|V|$ is the number of remaining AIAP candidates. AECI and AECP are standardized annually across the eligible country sample.
\end{tablenotes}
\end{threeparttable}
\end{table}

Japan leads AECI despite specializing in fewer listed goods than China, reflecting the higher average complexity of its specialized portfolio (Table~\ref{tab:country_portfolios}). AECI also differs from bundle-level trade specialization: the United States has a bundle RCA of 0.858 in 2023 but ranks fourth in AECI.

Across the 162 valid economies, AECI correlates with whole-space ECI at Pearson $r=0.744$ and Spearman $r_s=0.822$. Raw AECI, $F$, correlates with the number of AI-enabling specializations at $r=0.997$. Thus, although complexity weights can affect individual rankings, portfolio breadth accounts for much of the cross-country variation in this specification.

\subsection{Adjacent opportunities and product-space structure}
\label{sec:opportunities}
Remaining opportunities do not reproduce the ordering of current portfolios. China has the highest AECP, followed by Italy and Germany, whereas Japan leads AECI (Table~\ref{tab:country_portfolios}). AECI and AECP nevertheless correlate positively across the 162 valid economies in 2023 ($r=0.806$; $r_s=0.892$). This is a comparison of related measures sharing specialization data and complexity weights, not independent validation.

AIAP retains the product-level detail summarized by AECP. For South Korea in 2023, optical instruments for semiconductor inspection (HS 903141) are the highest-ranked remaining good by $\rho q$ ($\rho=0.257$, $q=0.825$). For Vietnam, memory integrated circuits (HS 854232) rank highest ($\rho=0.163$, $q=0.756$). These scores order candidates by the product of relatedness and complexity, not an estimated probability of future entry.

In the descriptive HS6 backbone, the 103 AI-enabling goods are distributed across 27 of 71 Louvain communities. Their induced subgraph contains 53 retained links and 46 isolated nodes. Isolation here means no retained AI-to-AI backbone edge, not an absence of connections to non-AI goods or in the full proximity matrix. Under this descriptive network's $\RCA\geq1$ convention, mean 2023 proximity is 0.212 among AI-enabling goods and 0.138 among non-AI goods. Higher average within-set proximity therefore coexists with dispersion across several backbone communities. These statistics describe the specified network representation; they do not identify the contribution of particular industries to subsequent diversification.

\subsection{Initial potential and subsequent change}
\label{sec:changes}
Table~\ref{tab:potential_change} reports the coefficient on baseline raw potential in Eq.~\eqref{eq:change_regression}. Each outcome has 141 usable economies after the period and data-availability requirements, compared with 162 in the 2023 descriptive comparisons.

\begin{table}[H]
\centering
\begin{threeparttable}
\caption{Association of baseline raw potential with subsequent outcome changes}
\label{tab:potential_change}
\small
\setlength{\tabcolsep}{6pt}
\begin{tabular*}{\textwidth}{@{\extracolsep{\fill}}lrrr@{}}
\toprule
Outcome change
& \shortstack{Coefficient on\\$\bar G_{0}$}
& HC1 SE
& $p$-value \\
\midrule
AI-enabling export share (0--1) & 0.1596  & 0.0625  & 0.0106 \\
AECI (standardized)             & 0.8645  & 0.8211  & 0.2924 \\
AI specialization count ($D^{\AI}$) & 16.4811 & 10.9136 & 0.1310 \\
\bottomrule
\end{tabular*}
\begin{tablenotes}[flushleft]
\footnotesize
\item Notes: One observation per economy; $N=141$ in each regression. The dependent variable is the difference between the 2019--2023 and 2007--2011 period means. Each model includes an intercept, baseline full-space ECI, the baseline mean of annual log real GDP per capita, and the baseline value of the corresponding outcome. The explanatory variable is raw $G$, not standardized AECP.
\end{tablenotes}
\end{threeparttable}
\end{table}

Higher initial raw potential is positively associated with the subsequent change in the AI-enabling export share ($\beta=0.1596$, $p=0.0106$). The coefficients for changes in standardized AECI and the specialization count are also positive, but are not statistically distinguishable from zero at the 5\% level ($p=0.2924$ and $p=0.1310$, respectively).

The association is therefore documented for export composition, but not established for every outcome in this specification. A change in export share need not coincide with entry into additional specialized products or an improvement in relative AECI. The estimates are in-sample conditional associations and do not identify causal effects, entry probabilities, or out-of-sample predictive performance.

\section{Discussion}
\label{sec:discussion}
The results show how a domain-focused economic complexity framework distinguishes aspects of AI-enabling trade that aggregate output measures combine. Japan's position relative to China in Table~\ref{tab:country_portfolios} illustrates the interaction of portfolio breadth and product complexity. The United States' high AECI despite a bundle RCA below one distinguishes a broad complexity-weighted portfolio from aggregate relative specialization in the bundle. These are differences among trade-based measures, not evidence that AECI outperforms publication or patent indicators in measuring national AI capability.

The positive correlation with whole-space ECI links the domain-focused ranking to broader export specialization, but is not external validation because the measures share trade inputs. The same results qualify the contribution of complexity weighting. The near-unit correlation between raw AECI and specialization counts (Section~\ref{sec:portfolios}) indicates that breadth dominates the observed cross-country variation. Complexity weights can alter particular rankings, as in the Japan--China comparison, without providing a large independent source of variation. AECI should therefore be interpreted together with its specialization count and mean-complexity components, rather than as evidence of a separately validated capability dimension.

Current portfolios and remaining opportunities also need to be distinguished. The different AECI and AECP leaders, and the different leading AIAP candidates for South Korea and Vietnam, show why an existing-portfolio ranking cannot substitute for a candidate-level assessment. The distribution of AI-enabling goods across multiple backbone communities is consistent with evaluating these candidates in the wider product space rather than treating them as a self-contained cluster. However, this description does not establish a capability-transfer mechanism or quantify the role of individual non-AI industries. AECP remains an average over each country's remaining candidates, not their number, aggregate value, or probability of entry.

The longitudinal results provide a limited additional observation. Initial raw potential is associated with subsequent changes in the AI-enabling export share, but the estimates do not establish corresponding associations for changes in AECI or specialization counts (Table~\ref{tab:potential_change}). This supports reporting an outcome-specific association with export composition, not a general validation of future capability growth. The non-significant coefficients do not establish that no relationship exists, and none of the regressions identifies a causal effect or tests out-of-sample prediction.

Interpretation remains bounded by the data and measurement design. Gross exports may include re-exports and imported inputs, while the listed categories include non-AI uses; they do not identify domestic AI-specific value added. Annual normalization and changing eligibility make AECI changes relative rather than absolute, and the retrospective product list does not identify historical AI demand. The measures also omit model performance, software and services, access to computing resources, and adoption. The study therefore offers a defined trade-based assessment, descriptive comparisons, and preliminary conditional associations within a generalizable framework. External validation and robustness to alternative classifications and complexity estimators remain separate tasks, not established properties of this version.

\section*{Data Availability}
The results are publicly available through the \emph{AI economic complexity} tab of the Economic Complexity Explorer:\\[-1pt]
\url{https://dawoon-jeong0523.github.io/EC_dashboard/dashboard.html}.
The dashboard presents country-year indicators, AIAP candidates, product-space views, and project figures. This note documents the product-only 2007--2023 results; Tables~\ref{tab:country_portfolios} and~\ref{tab:potential_change} summarize the country portfolios and preliminary longitudinal associations. The AI results use unscreened input matrices with the reporting eligibility rule in Section~\ref{sec:framework}, unlike the dashboard's separately screened datasets. Original trade and classification sources are BACI and the WTO \citep{gaulier2010,wto2025}; the dashboard is a dissemination site for derived results, not a complete archive of the raw input data.

\end{document}